# Machine Learning to Analyze Images of Shocked Materials for Precise and Accurate Measurements

Leora Dresselhaus-Cooper,[a,b] Marylesa Howard,[c] Margaret C. Hock,[c] B. T. Meehan,[c] Kyle Ramos,[d] Cindy Bolme,[d] Richard L. Sandberg,[d] Keith A. Nelson[a,b]

[a] *Department of Chemistry, Massachusetts Institute of Technology, Cambridge, MA, 02139, USA*

[b] *Institute for Soldier Nanotechnology, Massachusetts Institute of Technology, Cambridge, MA, 02139, USA*

[c] *National Security Technologies, North Las Vegas, NV, 89030, USA*

[d] *Los Alamos National Laboratories, Los Alamos, NM, 87545, USA*

A supervised machine learning algorithm, called locally adaptive discriminant analysis (LADA), has been developed to locate boundaries between identifiable image features that have varying intensities. LADA is an adaptation of image segmentation, which includes techniques that find the positions of image features (classes) using statistical intensity distributions for each class in the image. In order to place a pixel in the proper class, LADA considers the intensity at that pixel and the distribution of intensities in local (nearby) pixels. This paper presents the use of LADA to provide, with statistical uncertainties, the positions and shapes of features within ultrafast images of shock waves. We demonstrate the ability to locate image features including crystals, density changes associated with shock waves, and material jetting caused by shock waves. This algorithm can analyze images that exhibit a wide range of physical phenomena because it does not rely on comparison to a model. LADA enables analysis of images from shock physics with statistical rigor independent of underlying models or simulations.

## I. INTRODUCTION

Images that are acquired to characterize spatial variation in materials may show a wealth of information that cannot be obtained in other ways. Careful analysis of large datasets is often required for precise determination of material variability with meaningful uncertainties. In many settings, image acquisition is made difficult by factors including scattering of light, inhomogeneity of the light field that is used, small changes in material refractive index (i.e. low contrast), and other challenges. Interpretation of the resulting images can be strongly enhanced with advanced image processing approaches.

Image analysis algorithms have been widely developed for many different fields including satellite photography[1] and biological imaging.[2,3] These algorithms have used combinations of machine learning and artificial intelligence to extract precise locations for different types of objects. Image segmentation (also called image classification) has been used extensively to this end. Image segmentation is conducted using machine learning algorithms that separate (segment or classify) an image into different regions (classes). Supervised segmentation methods require the user to define a set of classes that correspond to features of interest (typically based on user insight about the image content) in the image.[4] Several fields of science have already adopted variations on segmentation to appropriately analyze images. Satellite tomography uses a partial differential equation (PDE) approach to locate sharp changes in the intensity gradients caused by the edges of trees, houses, rivers, etc. in order to identify image features without user-provided information (unsupervised).[1] Disease diagnostics use segmentation on microscopy images of samples from patients to separate benign cells from invasive bacteria in order to quickly determine if a patient has a bacterial illness.[2,3]

Images of shock waves and other high strain rate deformations are often difficult to interpret due to irreproducibility of successive measurements, unpredictable content in the images including spatially variable material responses and debris, spatial variations of the imaging light, and other causes. Because shock waves are high amplitude compression waves that move materials far from equilibrium, they usually cause irreversible damage such that each measurement must be made on a different sample or sample region, increasing variability from one measurement to the next. Reduction of shot-to-shot noise has presented challenges in engineering detectors with sufficiently high frame rates to obtain interpretable images with sufficient time resolution to measure shock-induced dynamics.[5–8] A single image recorded at a specified time may capture a substantial region that has been traversed by a rapidly propagating shock wave (on the order of kilometers per second), revealing very wide-ranging features. [9–12] Shock-induced transformations can include defect formation, fracture, phase transitions and chemical reactions whose spatial locations and extents in an image can reveal material properties and dynamics.[5,10,13,14] The combination of ultrafast imaging methods that capture spatially varying material responses with advanced image processing algorithms that reliably define the different regions under observation can further our understanding of complex shock-induced responses.[5,10,13,15–21]

In the field of shock physics, many images are manually segmented to extract material features from images.  Most image analysis uses Fourier filtering, which suppresses high-frequency noise, but also diminishes the high-frequency features inherent to shock waves.[22,23] Other image analysis algorithms rely on comparison of image features to theoretical models from fluid dynamics that describe the shock.[24,25] These algorithms can only be used on images collected for shocks that have well-characterized spatial dynamics. Images of shock waves or shock-induced responses that are not sufficiently well characterized to compare to a computational model require a general image analysis tool. Pribosek recently published a segmentation algorithm to perform unbiased image analysis using a PDE approach, which targets low-contrast images. This method has not been widely used within the field, as it is not optimized for high-noise images.[26] Established image segmentation algorithms cannot give accurate results due to high noise content in the images and large overlap in pixel intensities between the different classes, i.e. the different shocked states.

In this paper, we present the application of a recently developed image segmentation algorithm[27] to the analysis of images of shocked materials. Conventional segmentation methods were attempted unsuccessfully for some of the images, as described in the Supplemental Information. The algorithm is unique within machine learning in its ability to segment images with high noise and low variation between classes. For the first image from a series collected from converging shocks in water[28], we explain in detail the user-provided information and the algorithm processing routine through which a shock position is determined. We further illustrate the process for each of the subsequent images of the series. Shock positions with uncertainties are extracted from the set of images, and are used to calculate the presumed shock velocities with their corresponding uncertainties for the entire series. To show the wide applicability of this algorithm for images that are difficult to analyze, we present two additional images with many classes and several types of aberrations.[18] The analysis demonstrates the algorithm's ability to analyze different types of images in order to make precise and accurate measurements.

## II. ANALYSIS OF CONVERGING SHOCK WAVES IN WATER WITH SEGMENTATION

We describe statistical image segmentation that locates boundaries by sorting pixels from images into specific features (classes) based on the pixel intensity distributions. Supervised variants of statistical segmentation, like LADA, begin with the user providing a map to define the regions that are most characteristic of each class. LADA accounts for variation of intensities within a class by using both the intensity of each pixel and the distribution of intensities among surrounding pixels to sort the pixels into appropriate classes. The size of the surrounding region defines "local" for the algorithm, and can be set by the user with parameters $d$ and $n$, based on the distance scale over which significant intensity changes are observed. In contrast, as described in the Supplementary Information, some global methods create a histogram with the intensities of all of the pixels defined by sparse training data for each class.[4] The intensity distributions created from the well-defined regions of the image are then used to assign each pixel in the image to the class which best defines its intensity. Intensity variation within classes, or overlapping distributions for the intensities of different classes can cause the algorithm to misidentify large regions of the image.

To describe the mathematical process performed in LADA, we present its analysis of the first image from a sequence showing focusing shock wave propagation in water (Figure 1a). Images displaying real-time convergence of shock waves in this work were collected with a recently developed shock technique described fully elsewhere.[10,29,30] These laser-driven shocks were generated by the interaction of a ring-shaped beam with a thin soft sample. Placement of the sample between two thick glass wafers was predicted to confine the shock wave within the sample layer, allowing for imaging to capture the wave propagation.[10] Multi-frame imaging results similar to the ones shown in this work (using LADA) have indicated that further wave interactions may be present, suggesting possible multi-wave dynamics, as are being further investigated at this time.[28]

In our analysis of the strongest features evident in images of converging shocks, we demonstrate the use of analysis of variance (ANOVA) and maximum likelihood estimator (MLE) p-value maps to precisely calculate the boundary location uncertainties from the segmentation. The LADA measurements of shock position are then compiled from the entire sequence and are used to calculate the shock velocities and the uncertainties are propagated through the calculations. The results demonstrate the utility of this segmentation method, as quantitative determination of the shock properties from the images is difficult without the precision afforded by LADA.

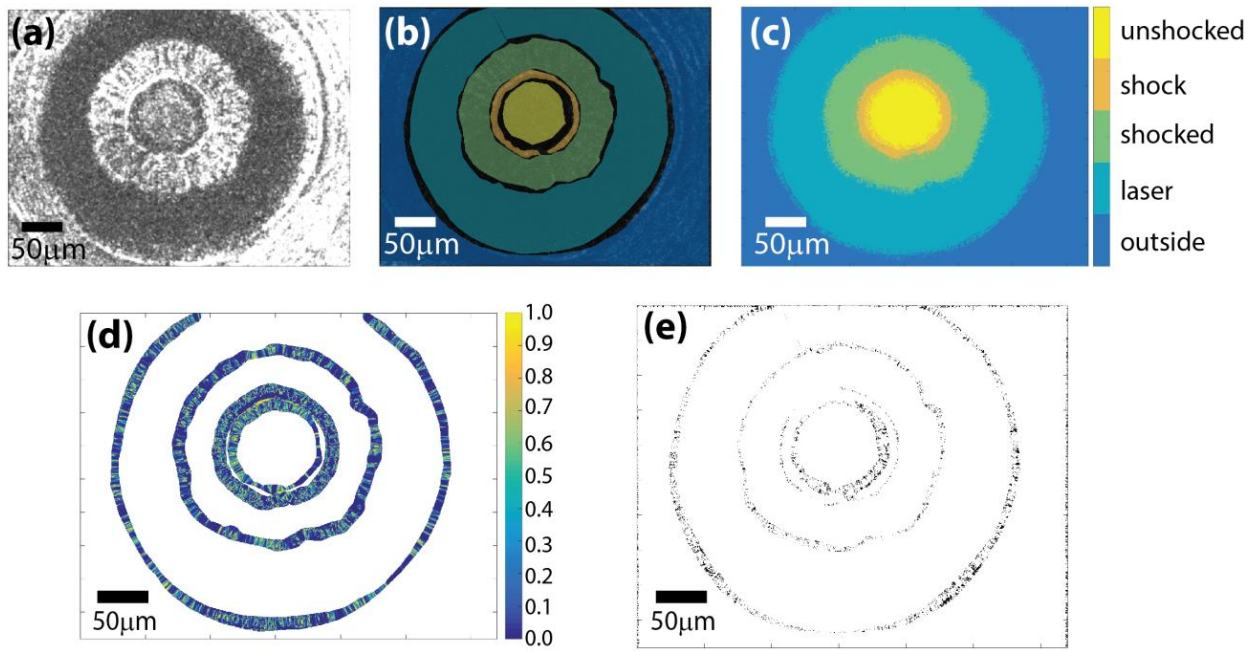

**Figure 1. (a)** The first of six ultrafast images measuring converging shocks travelling in a multi-layered target (water between glass). **(b)** The training data form a user-specified map of *a priori*, known regions for the different classes in the image. Each class is represented by a distinct color. The dark regions correspond to user-unspecified regions to be segmented by the program. **(c)** The segmented image ($d = 25$, $n = 5$) showing the locations for each class. The white regions are an added class that corresponds to pixels that did not have any local training data when the image was restricted by parameter $d$. **(d)** The ANOVA p-value map of the segmented pixels. **(e)** The maximum likelihood estimator (MLE) p-value map for the segmented image, thresholded at 5%. For this image, the $1280 \times 960$ pixels are either black or white based on the thresholding, as mentioned in the text.

Use of LADA first requires the user to provide training data, which gives *a priori* knowledge of the image features of interest to the algorithm. Training data form a map that can be overlaid on top of the image to show the approximate class locations in the image (Figure 1b). As a local algorithm, LADA requires the training data to provide regions of the image that are representative of both characteristic pixel intensities and positions for each class. LADA best analyzes images consisting mainly of classes that may be reliably definable by the user, and whose unknown regions are localized around the boundaries between classes. LADA analysis provides the locations of the boundaries.

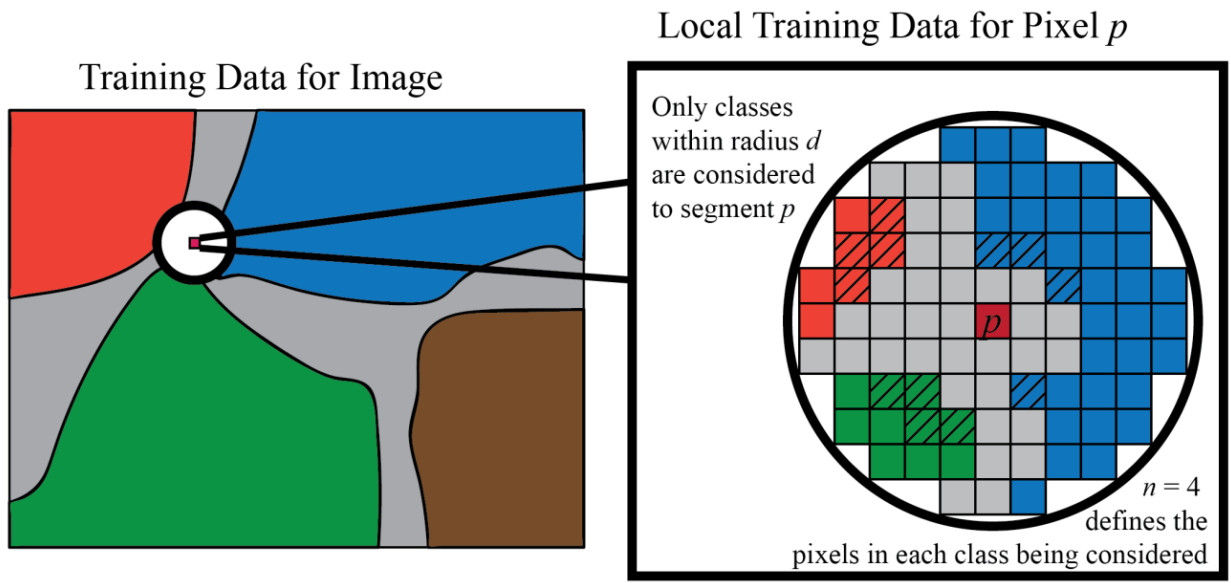

**Figure 2.** An illustration of the method by which LADA selects local training data for the image. Training data for an entire image (left) has four classes, shown in orange, green, blue and brown. The undefined region for the segmentation is shown in grey. For the red pixel $p$, a circle of radius $d$ defines the set of pixels considered for the local training data (right). Within the local training data, the $n$ nearest pixels used by the method are highlighted with stripes.

As shown with a cartoon in Figure 2, LADA also requires the user to provide two parameters, *d* and *n*, which define "local" for the algorithm. The value of *d* sets the distance within which the distribution of pixel intensities may be considered in assigning the class for a given pixel. Large values of *d* allow all of the defined classes to be considered for every pixel, while low values of *d* limit the number of classes available for comparison to those within the radius *d* around the pixel. Intensity histograms are assembled using the *n* nearest pixels in every class within the radius *d*. If there is a tie between pixels that are the same distance from *p*, the *n* points are selected using lexicographical ordering. High input values for both of these parameters bias the segmented output toward a global result. The optimal value of *d* will balance selecting enough local classes with avoiding the inclusion of training data that do not represent the local behavior around that pixel. That helps ensure that pixels that belong to one class are not misassigned to another class with a similar distribution. The optimal value of *n* typically corresponds to the number of pixels over which the local intensity variation within the class is homogenous. This helps avoid unwanted fluctuations within the histograms that are constructed. Pixels for which the radius *d* includes only one region are auto-assigned to that region. For a class to be considered local, we require at least three training pixels within radius *d* to allow us to gather its intensity information statistically. If a pixel does not have at least three training pixels of any local user-defined class, it is assigned to a *bonus class*, indicating uncertainty due to sparse training data. To summarize, the value of *d* can be set to affect the number of classes being considered, while the value of *n* determines the number of pixels compiled into each histogram.

Once the user inputs have been collected, the algorithm locally segments the pixels into classes. The algorithm progresses through all of the pixels within the image, to assign each one to the class that is most probable, based on both its intensity and position. Uncertainty in class assignment may still be present due to incorrect training data, insufficient training data, or random noise caused by the camera or experiment that is manifested in the image. For each pixel, the algorithm crops the image within a radius *d* and for each class specified in the training data creates a histogram using up to *n* pixels. If the value of *n* exceeds the number of pixels in a given class within the cropped image, the histogram is formed by only the pixels that are within the cropped image. Each histogram is fitted to a Gaussian curve. The intensity of the pixel is then compared against the Gaussian distributions for each local class, and is assigned to the class with the highest probability for that intensity. The final segmentation of Figure 1a is given in Figure 1c, using $d = 20$ and $n = 10$.

Segmentation uncertainty is measured for each pixel in two different ways, which cover the two most common causes of imprecise segmentation. Uncertainty can arise when multiple local classes have broadly overlapping intensity histograms, in which case some pixels with similar intensities may fall into

~ 6 ~

more than one class and there is a risk of misidentification. The uncertainty caused by overlapping distributions from local classes is characterized by analysis of variance (ANOVA),[4] as shown in Figure 1d. Alternatively, uncertainty can arise when pixel intensities yield a low probability of belonging to any of the local classes specified in the training data. This uncertainty is measured by compiling an MLE uncertainty map[1,4] ($\alpha = 0.05$) to describe the p-values that in our work are thresholded at 5% (Figure 1e).

ANOVA measures the extent of overlap among all of the local histograms there were constructed for the assignment of a given pixel. The p-values for ANOVA range from 0 to 1, with ascending values indicating a higher degree of overlap among the classes defined by $d$ and $n$ to be local within the training data. Pixels with high ANOVA values have poor differentiability among the available local classes. ANOVA maps that show high overlap among classes can be indicative of either non-optimized parameters for the user-inputs or regions of the image that cannot be precisely segmented. In determining the optimal parameters for LADA segmentation, ANOVA values can often be reduced by modifying the training data and the values of $d$ and $n$. High ANOVA values often indicate that the image requires smaller values of $d$ and $n$ to correctly define the local classes according to the length scale over which intensity varies within a class. Inaccurate training data that include pixels from adjacent image features result in a local histogram for the class that is not representative of the *a priori* knowledge from the user. In those cases, two overlapping Gaussian fits of the histograms can show unfitted distributions that are both bimodal, representing one mode from each class. Low local contrast between image features covering a part of the image that are not well measured by LADA can be seen as regions for which the ANOVA values cannot be reduced by changing the user-inputs. The white regions in Figure 1d are areas that were assigned by the user and for which no ANOVA value was computed.

Regions of the image that are not well-defined by any of the local classes are shown in Figure 1e with an MLE p-value map thresholded at $\alpha = 0.05$.[31] To compute this error map, p-values are calculated for each pixel based on the probability of that pixel being within the selected class. Calculated p-values are between 0 and 1, with higher values corresponding to high confidence for the segmentation. The uncertainty map in Figure 1e is thresholded to only have a black pixel corresponding to a p-value that is ≤0.05, or for any pixel that was placed into a bonus class. Figure 1e shows a random distribution of dark pixels throughout the image that corresponds to noise in the original image. A high density of dark pixels – located at a boundary between classes – indicates the uncertainty from the segmentation between the local classes for that region. The map of regions with low-confidence segmented pixels best describes the uncertainty in measurements of the boundaries between the adjacent classes and is used for propagation of error in subsequent calculations.

Appropriate results for the images in this paper were acquired by iteratively running the algorithm to optimize the user-input parameters consistent with *a priori* knowledge of the experimental system and the underlying science. All images shown in this paper were initially segmented using sparser training data than are shown here and high values for *d* and *n* to allow the algorithm to discern boundaries that are difficult to see by eye. After each iteration, we refined the training data and the local parameters based on the boundaries and sharp features identified by the method. The procedure ended when there were no significant changes between iterations and our *a priori* knowledge does not suggest any further refinements to the training data or the local parameters.

Our primary interest with the LADA results was in locating the position of the leading edge of the dark feature we interpreted as the shock which, in Figure 1, is the innermost boundary. We defined the two regions on either side of this boundary, "shock" and "unshocked," as uniform arrays of 1 and 0 values, respectively. The positions of each pixel of the shock front was measured by taking the gradient of the segmented image to locate the edges between the classes.[32] The resulting points were input into a circle fitting algorithm[33] to define the center of the shock front. To measure the shock radius, we transformed the segmentation result into polar coordinates, and divided the shock ring into 60 azimuthal sections to make the measurement statistically. The position corresponding to a maximum value for the second derivative of the segmentation result in each azimuthal section was used to measure a distribution of shock radii for each image. The mean values are shown in Figure 4, with two times the standard deviation (95% certainty) shown by the error bars. The uncertainty that the algorithm had in locating the boundary was obtained from the MLE plot. That uncertainty was collected using 60 azimuthal slices to measure the position with the highest weighted uncertainty around that boundary. The resulting distributions then gave the mean and standard deviations for the position of highest uncertainty for each azimuthal slice. The two standard deviations about the mean position of highest uncertainty was extracted from the distribution for each image, corresponding to the positions of highest LADA uncertainty about the shock front.

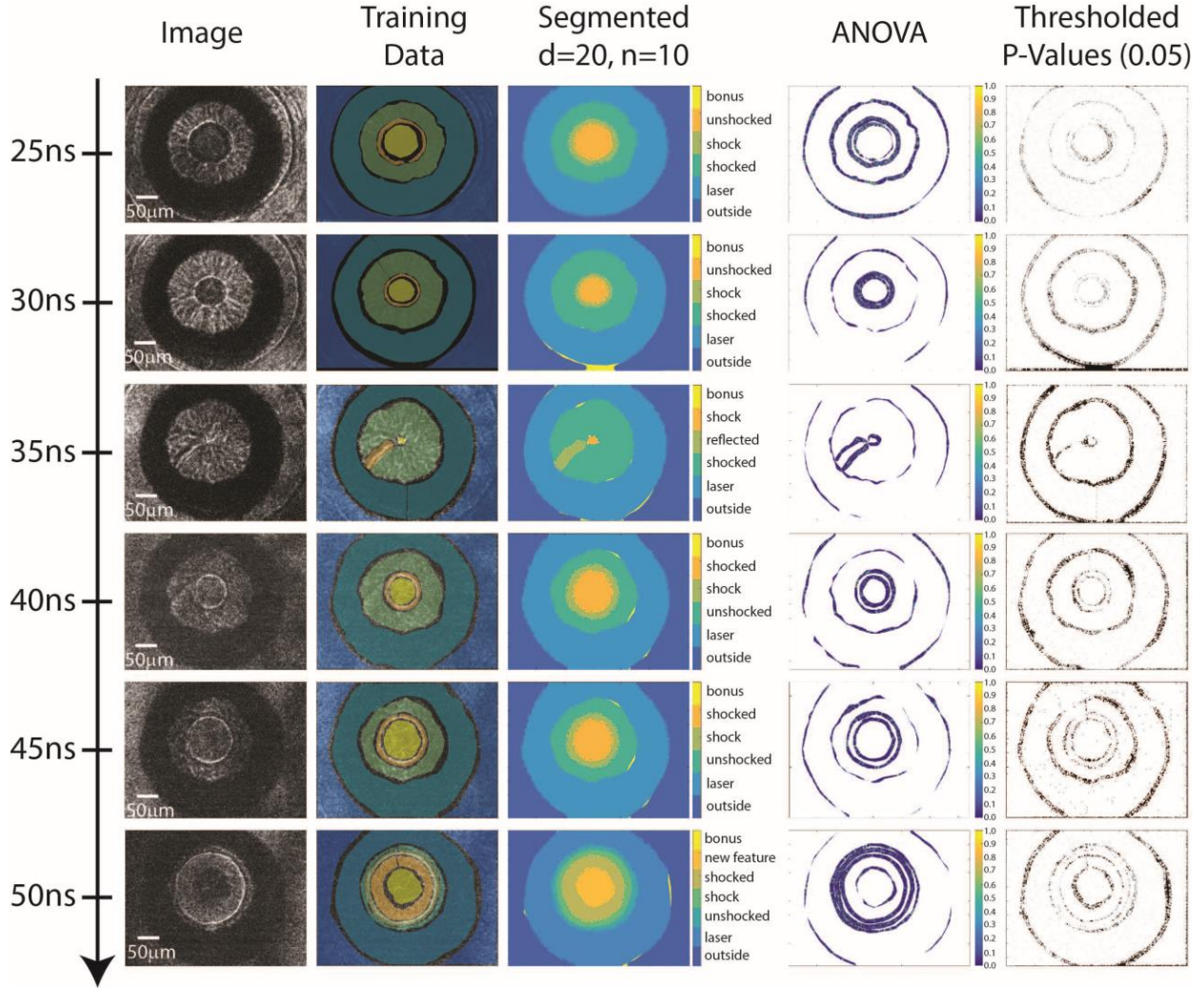

**Figure 3.** LADA segmentation of the entire image sequence for a converging shock wave in water, using parameters $d = 20$ and $n = 10$. For each of the images in the sequence, a bonus class was generated due to uncertainty in assigning sufficient training data for that region of the image. Initially, a shock wave converges into a central region, then it diverges outwards. At the point of convergence, geometric instabilities create a Mach reflection,[34] observed at 35 ns. After the shock diverges, additional features become evident beginning at 50 ns, which continue after the end of this image sequence.[10]

The LADA analysis of all of the images shown in Figure 3 was used to determine the shock position for all six images in the sequence. The shock positions, calculated as described above, were located for all time delays, and the average shock velocities were calculated with

$$U_s(t_n) = \frac{r_n - r_{n-1}}{t_n - t_{n-1}} \tag{1}$$

for each pair of successive images, based on the measured time delay $t_n$ and shock position $r_n$ measured from segmentation and subsequent analysis. The average shock velocity $U_S$ between each pair of images was calculated for all 60 azimuthal sections around the shock ring, to create a distribution of shock velocities. The plots in Figure 4 show the shock positions and shock velocities from the image sequence shown in Figure 3. For all of the data, the points and their error bars show the statistical values for position and velocity for each image, collected from the segmentation result. The positions and velocities that had the highest uncertainty in the segmentation are shown in the shaded gray and blue regions, respectively, behind the curves. For all of these values, two times the standard deviations (2σ) are shown, corresponding to the intervals of 95% confidence, consistent with the MLE confidence interval.

The difference between the error bars and the shaded uncertainty bands highlight the unique separation of uncertainty that LADA provides. The azimuthal statistics for the shock positions and velocities from the segmentation result allow us to see variation in the shock front position. Variation in the position is caused by ellipticity in the shock ring (systematic error), and mis-assigned pixels (random and systematic error). Statistical variation in the velocity values include systematic error from variation in the shock velocity around the ring, which is often seen in converging shock waves with geometric distortions.[35] The standard deviations show the extent of variation, but cannot determine the asymmetry of uncertainty, which can cause the mean value to be skewed. Deviation between the bands of MLE uncertainty and the error bars demonstrate results for which the mean value calculated from the segmentation result is skewed by uncertainty from the method's ability to segment that region of the image.

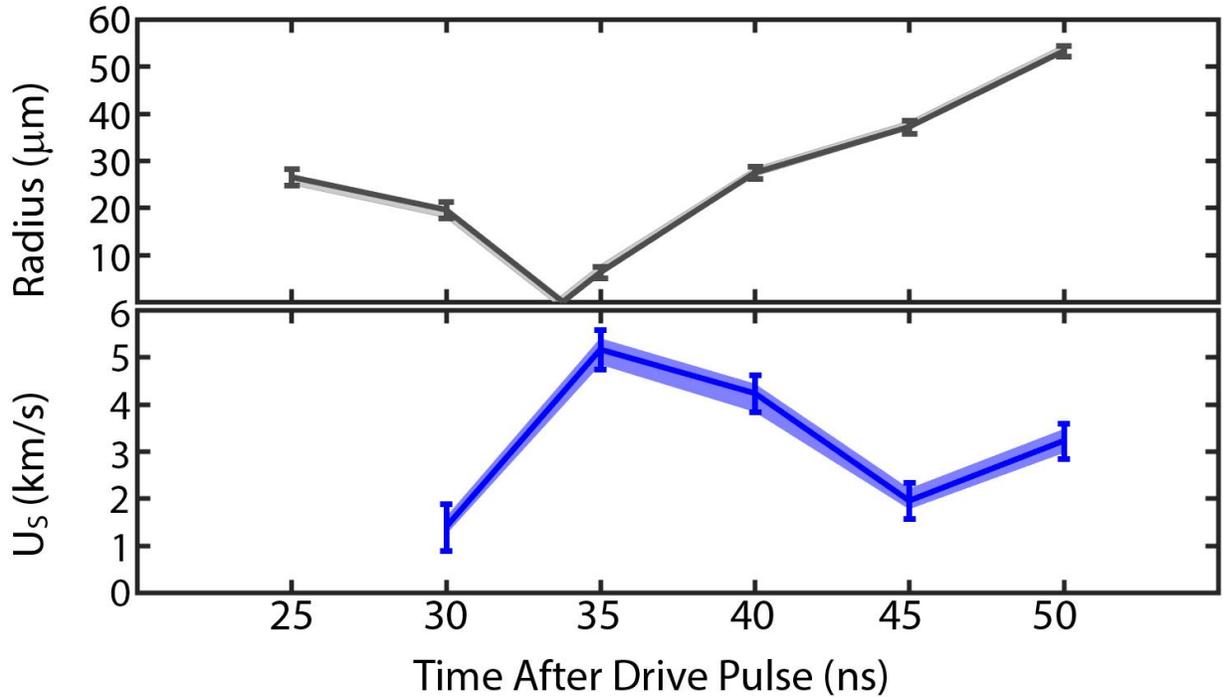

**Figure 4.** Measurements of shock positions and calculated velocities for the entire series of images shown in Figure 3. The measurements were done using the LADA analysis with $d = 20$ and $n = 10$ for each image in the sequence (analysis shown in Figure 3). The points and error bars show the mean and $2\sigma$ values for the shock position and velocities, calculated statistically from azimuthal slices around the shock ring. Shaded gray and blue regions behind the curves show the positions and corresponding velocities for which the algorithm had ≤5% confidence, as derived from the MLE plots.

The temporal separation between the images was measured using an oscilloscope with 250 ps resolution, giving a separation of $5 \pm 0.25$ ns between frames. Temporal uncertainty was not included for this calculation, as the uncertainty in velocity is dominated by the much larger uncertainty in position. For experiments that do not have well-defined temporal resolution, an RMS average may be taken to account for the temporal and spatial uncertainty. New understanding of this experiment suggests that multi-wave interactions preclude the use of the principal Hugoniot to infer shock pressures from the velocity of the shock-induced image features. For other systems, the velocity measurements from LADA can be input to the equation of state in order to infer the shock pressure and uncertainty at each time.

# III. EXTENDING LADA TO FURTHER COMPLICATED SHOCK IMAGES

In the water image sequence above, we have explained the LADA methodology for the segmentation of converging waves in targets containing water for our experimental geometry. This analysis demonstrates the utility of LADA in finding the boundaries between the shock wave and unshocked material in order to measure shock position and velocity. We now demonstrate LADA's ability to separate locally varying images for which the boundaries between classes are difficult to identify by eye. Figure 5a shows an image captured with a 400-nm laser pulse using shadowgraphy (second derivative of density) of a converging shock wave transducing into an RDX crystal embedded in a polymer. Figure 6a shows an X-ray phase-contrast image of shock-induced void collapse causing jetting in acetaminophen. Subsequent figures in the respective sequences present the segmentation for the two different experiments. These images show the utility of the algorithm in extracting scientifically meaningful boundary measurements from the image segmentation afforded by LADA even with high intensity variation within individual classes.

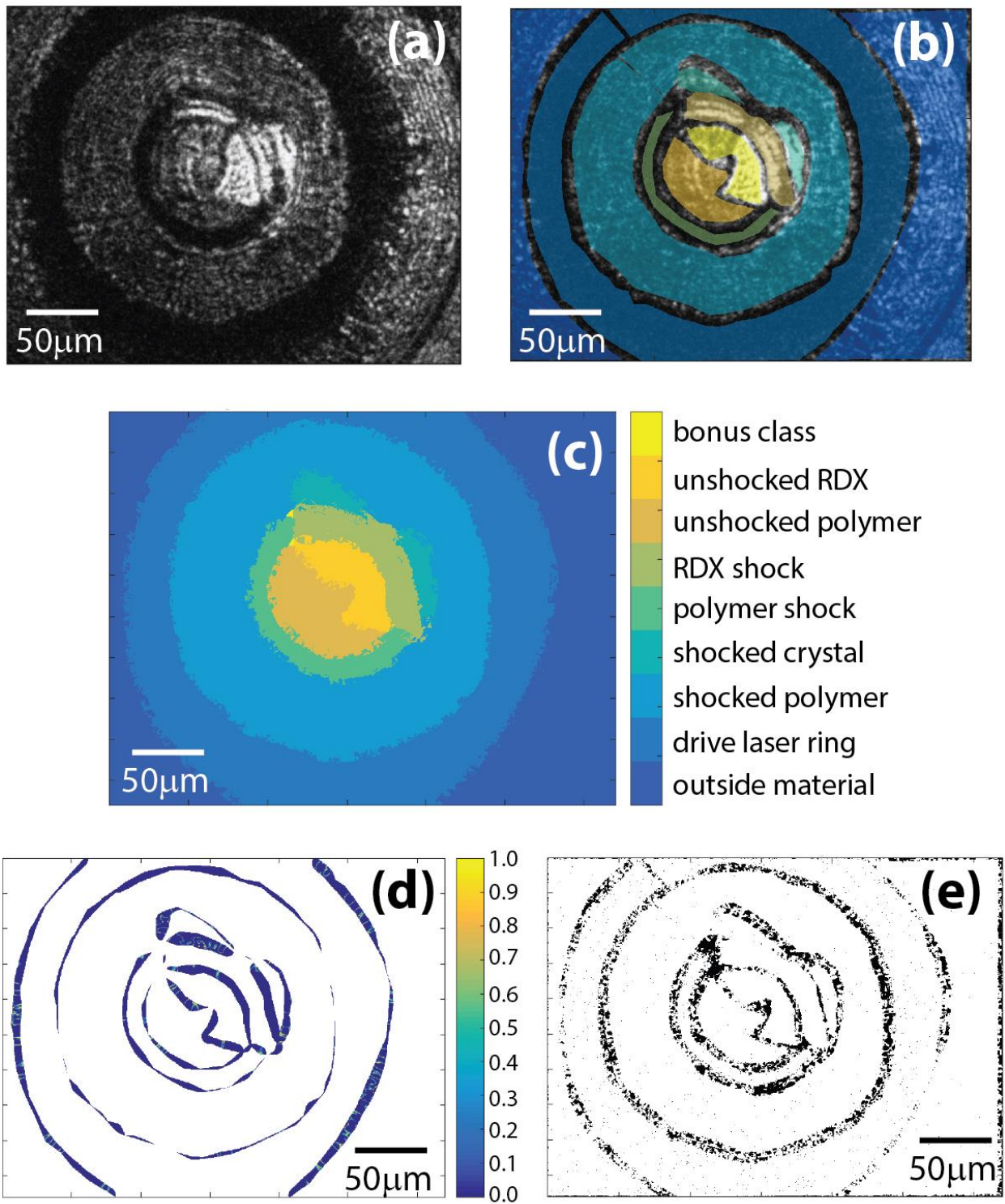

**Figure 5.** The inputs and outputs showing the segmentation of (**a**) an image with a shock feature that appears mostly in the polymer, but partly in a crystal of RDX. The training data (**b**) show that there are many classes, describing a complicated experimental geometry. The segmented result using $d = 20$, $n = $

15 **(c)** shows that the segmentation algorithm locates boundaries than the simple inspection. Error maps of **(d)** ANOVA and **(e)** MLE plots demonstrate that the image yields considerable uncertainty but can still be segmented reliably.

Figure 5 shows the segmentation of an image which we understood to show a shock travelling through our converging shock geometry that transduces from polymer into an RDX crystal upon convergence. This image has many more classes than were seen in Figure 1. The *a priori* classes based on the experimental conditions are: the laser ring from the drive beam, the shocked polystyrene, the shock wave in the polystyrene, the unshocked polystyrene, the shocked crystal, the shock wave in the crystal, the unshocked crystal, and the region outside of the drive laser ring. The high number of classes for this image and the density of the classes make the image difficult to separate manually. As seen from the segmentation in Figure 5, LADA was able to locate boundaries between classes that were difficult to identify visually. While the classes are difficult to distinguish visually, the ANOVA error is quite low, indicating that the classes are locally quite well defined. The MLE plot demonstrates that the noise within the images generates substantial uncertainties for many but not all of the boundaries. Despite high uncertainty in some regions, the statistical analysis of LADA measures boundaries in Figure 4 that are difficult to discern by eye. This analysis allows for measurement of the features and structural variation of the shape change of the feature we interpret as the shock wave after it transduces into the polymer.

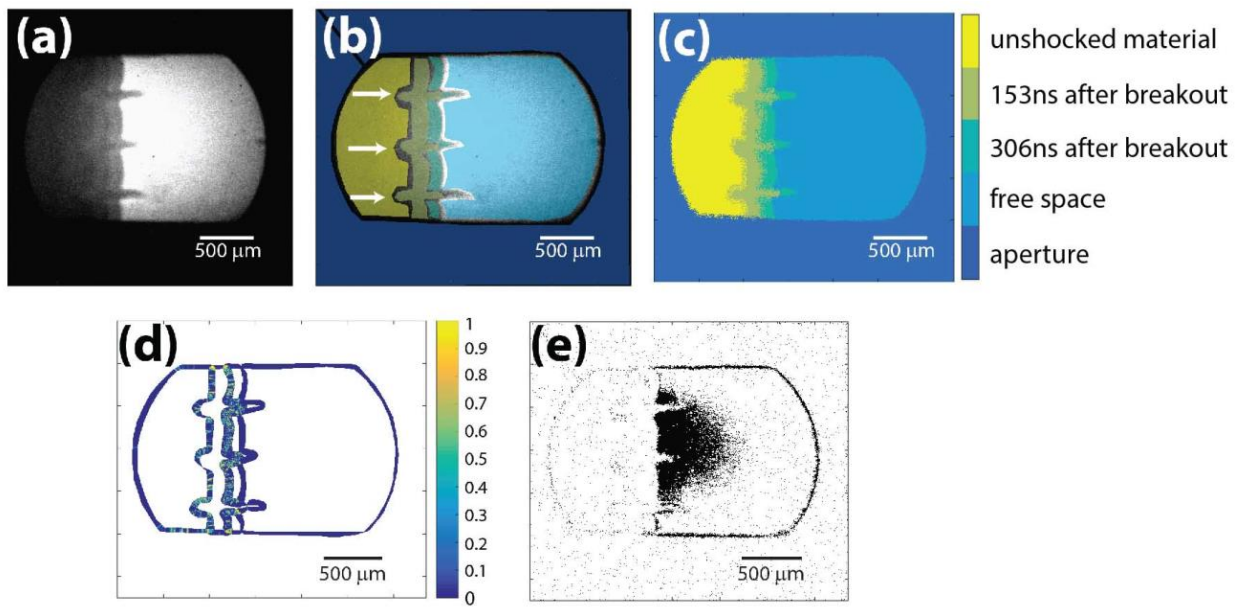

**Figure 6.** Analysis of an image **(a)** of void collapse and the resultant jetting, as caused by a planar shock in acetaminophen collected with X-ray phase contrast imaging. The training data **(b)** includes arrows to indicate the original positions of the voids. The training data and parameters $d = 24$, $n = 10$ were used to

~ 14 ~

segment the image **(c)** with high precision, despite the aberrations from ghosting and saturation of the detector (photo bleaching of the scintillator) in the experiment. The error maps of **(d)** ANOVA and **(e)** MLE demonstrate the high accuracy of the measurement, despite features that are hard to discern in the original image. Reproduced with permission from J. Phys. Conf. Ser. 500, (2014) p. 142028. Copyright IOP Publishing.

Figure 6 illustrates the use of LADA to analyze a phase contrast image from an 80-ps duration synchrotron X-ray pulse that shows a planar shock wave causing acetaminophen to jet as shocked voids collapse.[6] This image shows an aperture within which appears an image of the jetting material extending into free space. The three different intensity classes in Figure 6b-c arise from ghosting of images from a series (153 ns inter-frame) caused by the scintillator detector.[7] This ghosting causes Figure 5a to show three overlapping images with different intensities, each of which corresponds to an image produced by X-rays from adjacent synchrotron pulses during the jetting event. In this case, segmentation of the image can provide the temporal evolution of the jetting in the material.

Analysis of this image (and similar ones) has been difficult, as the unsegmented image (Figure 6a) has boundaries that are challenging to discern with statistically accurate uncertainty. The segmentation with LADA allows for precise location of boundaries from Figure 6a between the different times within the jetting with relative ease. The uncertainty shown in the ANOVA plot demonstrates that the classes are locally distinguishable for most of the image.

Figure 6e shows an MLE plot for the segmentation, which shows a case that is not seen in the previous examples. With the exception of the white class, boundaries between each class in the image show high confidence from the thresholded plot. There is a region of low p-values and high uncertainty which is located at the center of the image, in the white class. In this case, high uncertainty corresponds to a region which is defined by the user in the training data. The high uncertainty found from the white class is not an artifact, but is due to lack of local intensity variation in that class. The intensity distribution in the region of high uncertainty is extremely narrow, as the camera was saturated for those pixels. Saturated signal causes the distribution to narrow toward a delta function, which makes any small amount of variation in pixel intensity have extremely low overlap with the local distribution. Knowing which regions of an image gave a false signal due to saturated detectors was important to avoid misinterpretation of the science corresponding to the feature intensities. The same trend was observed in the aperture region, due to lack of variation within the class. Segmentation was completed successfully by artificially introducing noise to the lowest intensities to generate a statistical distribution in order to correctly identify the aperture class.

The application of LADA for quantitative analysis of ghosted images similar to Figure 6 has significant potential for use of recently developed pulse sequences in real-time imaging. These pulse sequences include the use of mode 324 at the Advanced Photon Source, which has 11.37 ns between pulses of 80ps duration.[36] Use of X-rays for single-shot multi-frame imaging has been limited because there current scintillators do not have sufficient conversion efficiency from X-ray to optical light or decay times from that conversion process. Attempts at multi-frame imaging with these slow detectors causes ghosting and saturation (photo bleaching) by subsequent images.[37] Improved methods for extracting data from sub-optically recorded images like the ones shown in this paper provide a possible route for data analysis using detectors near their temporal limits.

## IV.   CONCLUSIONS

The results presented here explain the use of LADA to extract quantitative information from images with rich information content about materials as they are shocked. With LADA, the computer uses *a priori* knowledge from the user to learn the approximate positions and local intensity distributions for each of the features of interest in the image. The algorithm then assigns all of the pixels in the image to the classes that best describes them. LADA outputs the segmented image with boundaries and their uncertainties determined. ANOVA uncertainty is used to optimize the user-input parameters and to identify regions that segmentation is not able to identify. An MLE p-value map is used to determine uncertainty in the local segmentation for each pixel, and is used to extract error bars from measurements made with LADA. We used LADA to determine the shock positions and velocities, including their uncertainties, for an image series showing a shock wave converging and then diverging in water. The generality of the algorithm is shown through the segmentations of complicated images with significant aberrations and variation of intensity across features.

Software for LADA is available through individual license. Please contact Marylesa Howard at the Nevada National Security Site: howardmm@nv.doe.gov

**Supplementary Material**

See supplementary material for the results of global segmentation using quadratic discriminant analysis, using the image and training data from Figure 1.

**Acknowledgements**

The authors would like to thank Aaron Luttman and Jing Kong for their suggestions and contributions to this work. Figure 6 has already been published by Ramos, et al. as cited in this work.[6]

The authors would like to acknowledge the Office of Naval Research for funding that supported two of the authors on from this work on grant numbers N00014-16-1-2090 and N00014-15-1-2694.

This manuscript has been authored in part by National Security Technologies, LLC, under Contract No. DE-AC52-06NA25946 with the U.S. Department of Energy, National Nuclear Security Administration, NA-10 Office of Defense Programs, and supported by the Site-Directed Research and Development Program. The United States Government retains and the publisher, by accepting the article for publication, acknowledges that the United States Government retains a non-exclusive, paid-up, irrevocable, world-wide license to publish or reproduce the published content of this manuscript, or allow others to do so, for United States Government purposes. The U.S. Department of Energy will provide public access to these results of federally sponsored research in accordance with the DOE Public Access Plan (http://energy.gov/downloads/doe-public-access-plan). DOE/NV/25946--3183

The authors would like to acknowledge support from the U.S. Department of Energy through LANL's Laboratory Directed Research and Development program. LANL is operated by Los Alamos National Security, LLC for the U.S. Department of Energy under contract DE-AC52-06NA25396.

# Supplemental Information

## GLOBAL SEGMENTATION OF IMAGE FROM FIGURE 1

The main text from the paper describes segmentation by a local algorithm that accounts for variation in image intensity across classes. Segmentation of the images shown in Figure 1-2 were attempted using established global segmentations, but the pixels were incorrectly sorted due to overlapping intensity distributions between classes. Figure S–1 demonstrates the unsuccessful segmentation of the image shown in Figure 1 using Quadratic Discriminant Analysis (QDA).[1]

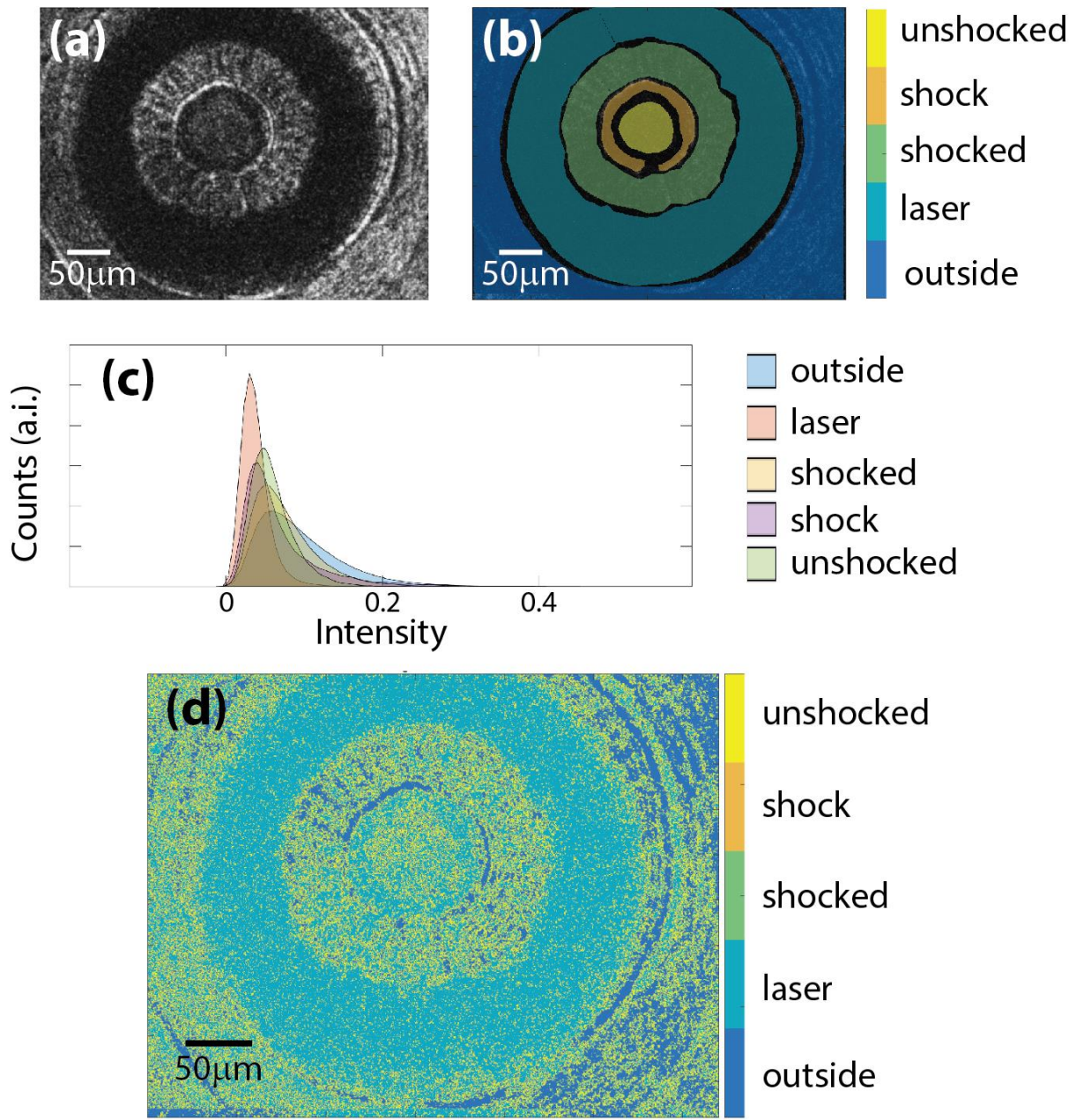

**Figure S–1.** The QDA segmentation of the same image analyzed in Figure 1 from the main paper using LADA. The original figure (a) is shown again, with the exact same training data (b) as was shown in Figure 1, and used for the calculations that followed. The global histograms corresponding to the fits from each class are shown (c) and the resulting global segmentation is presented in (d).

The QDA analysis presented in Figure S–1 used the same image and training data from the LADA analysis in Figure 1. As a global method, QDA compiled a histogram from all of the pixel intensities in the training data to create the Gaussian fit for the subsequent segmentation. The wide intensity variation within each class in Figure S–1 caused all of the histograms to overlap substantially, as shown in Figure S–1c. Overlapping histograms caused QDA to misidentify many pixels, as they often had high probability of belonging to more than one class. The main paper measures uncertainty from overlapping distributions using ANOVA p-values. Despite the global ANOVA uncertainty, Figure S–1a has visually discernable features that are not fully captured by QDA. In Figure S–1a, globally overlapping distributions arise from classes with similar intensities being spread out across the image. The image shown in Figure S–1a failed to be appropriately segmented by QDA because the presence of non-adjacent classes with similar intensities required the position of each pixel to be considered by the method. A thorough comparison of local and global histograms can be found in the seminal paper on LADA.[2]